%% file: paper.tex
\documentclass[conference]{IEEEtran}

\input{head}

\input{include}

\begin{document}

\title{High-Throughput Computing on \\ High-Performance Platforms: A Case Study}

\author{
    \IEEEauthorblockN{
        Danila Oleynik\IEEEauthorrefmark{1}\IEEEauthorrefmark{5},
        Sergey Panitkin\IEEEauthorrefmark{1}\IEEEauthorrefmark{4},
        Matteo Turilli\IEEEauthorrefmark{1}\IEEEauthorrefmark{2},\\
        Alessio Angius\IEEEauthorrefmark{2},
        Sarp Oral\IEEEauthorrefmark{6},
        Kaushik De\IEEEauthorrefmark{5},
        Alexei Klimentov\IEEEauthorrefmark{4},
        Jack C. Wells\IEEEauthorrefmark{7} and
        Shantenu Jha\IEEEauthorrefmark{2}\IEEEauthorrefmark{3}
    }
    \IEEEauthorblockA{
        \IEEEauthorrefmark{1}First authors, alphabetical order
    }
    \IEEEauthorblockA{
        \IEEEauthorrefmark{2}Department of Electric and Computer Engineering, Rutgers University, NJ, USA
    }
    \IEEEauthorblockA{
        \IEEEauthorrefmark{3}Computational Science Initiative, Brookhaven National Laboratory, NY, USA
    }
    \IEEEauthorblockA{
        \IEEEauthorrefmark{4}Physics Department, Brookhaven National Laboratory, NY, USA
    }
    \IEEEauthorblockA{
        \IEEEauthorrefmark{5}Physics Department, University of Texas Arlington, TX, USA
    }
    \IEEEauthorblockA{
        \IEEEauthorrefmark{6}Oak Ridge Leadership Computing Facility, Oak Ridge National Laboratory, TN, USA
        }
    \IEEEauthorblockA{
        \IEEEauthorrefmark{7}National Center for Computational Sciences, Oak Ridge National Laboratory, TN, USA
        }
}

\maketitle

\begin{abstract}
\input{abstract}
\end{abstract}

\section{Introduction}\label{sec:intro}
\input{introduction}

\section{PanDA Overview}\label{sec:panda_overview}
\input{PandaOverview}

\section{Deploying PanDA on Titan}\label{sec:panda_titan}
\input{PandaEvolution}

\section{Analysis and Discussion}\label{sec:analysis}
\input{overall}

\section{PANDA\@: The Next Generation Executor}\label{sec:panda_roadmap}
\input{nge}
\input{Experiments}

\section{Related Work}\label{sec:related}
\input{bckground}

\section{Conclusion}\label{sec:conclusion}
\input{conclusion}

\bibliographystyle{IEEEtran}
\bibliography{bibliography}

\end{document}

%% file: head.tex
\usepackage{graphicx}
\usepackage{amsmath}
\usepackage{amssymb}
\usepackage{color}
\usepackage{ifpdf}
\usepackage{float}
\usepackage[utf8]{inputenc}
\usepackage{multirow}
\usepackage{rotating}
\usepackage[caption=false]{subfig}

\usepackage{moresize}
\usepackage{url}
\usepackage{booktabs}
\usepackage{listings}
\usepackage{paralist}
\usepackage{wrapfig}
\usepackage{multirow}
\usepackage{ifpdf}
\usepackage{xspace}
\usepackage{keyval}
\usepackage{color}

\usepackage{comment}

\definecolor{listinggray}{gray}{0.95}
\definecolor{darkgray}{gray}{0.7}
\definecolor{commentgreen}{rgb}{0, 0.4, 0}
\definecolor{darkblue}{rgb}{0, 0, 0.4}
\definecolor{middleblue}{rgb}{0, 0, 0.7}
\definecolor{darkred}{rgb}{0.4, 0, 0}
\definecolor{brown}{rgb}{0.5, 0.5, 0}
\definecolor{dkgreen}{rgb}{0,0.5,0}
\definecolor{orange}{rgb}{1,.5,0}
\definecolor{dandelion}{cmyk}{0,0.29,0.84,0}

\usepackage[normalem]{ulem}
\makeatletter
\def\cyanuwave{\bgroup \markoverwith{\lower3.5\p@\hbox{\sixly \textcolor{cyan}{\char58}}}\ULon}
\def\reduwave{\bgroup \markoverwith{\lower3.5\p@\hbox{\sixly \textcolor{red}{\char58}}}\ULon}
\def\blueuwave{\bgroup \markoverwith{\lower3.5\p@\hbox{\sixly \textcolor{blue}{\char58}}}\ULon}
\font\sixly=lasy6 
\makeatother

\usepackage{enumitem}

%% file: include.tex
\newif\ifreview

\newif\ifdraft
\ifdraft
 \newcommand{\N}[1]{\textbf{NOTE: #1}\xspace}
 \newcommand{\jhanote}[1]{ {\textcolor{red} { ***SJ: #1 }}}
 \newcommand{\sergeynote}[1]{ {\textcolor{blue} { ***SP: #1 }}}
 \newcommand{\mtnote}[1]{ {\textcolor{orange} { ***MT: #1 }}}
 \newcommand{\jonnote}[1]{ {\textcolor{dkgreen} { ***JW: #1 }}}
 \newcommand{\aanote}[1]{ {\textcolor{dkgreen} { ***AA: #1 }}}
 \newcommand{\note}[1]{ {\textcolor{brown} { *** #1 }}}
 \newcommand{\sarp}[1]{{\color{red}\textit{sarp: {#1}}}}
\else
 \newcommand{\N}[1]{}
 \newcommand{\jhanote}[1]{}
 \newcommand{\sergeynote}[1]{}
 \newcommand{\mtnote}[1]{}
 \newcommand{\jonnote}[1]{}
 \newcommand{\sarp}[1]{}
 \newcommand{\aanote}[1]{}
 \newcommand{\note}[1]{}
\fi

\newif\ifdraft
\ifdraft
\newcommand{\terminology}[1]{ {\textcolor{red} {(Terminology used: \textbf{#1}) }}}

\newcommand{\alnote}[1]{ {\textcolor{blue} { ***andreL: #1 }}}
\newcommand{\amnote}[1]{ {\textcolor{blue} { ***andreM: #1 }}}
\newcommand{\smnote}[1]{ {\textcolor{brown} { ***sharath: #1 }}}
\newcommand{\pmnote}[1]{ {\textcolor{brown} { ***Pradeep: #1 }}}
\newcommand{\msnote}[1]{ {\textcolor{cyan} { ***mark: #1 }}}
\newcommand{\mrnote}[1]{ {\textcolor{purple} { ***melissa: #1 }}}
\else
\newcommand{\onote}[1]{}
\newcommand{\terminology}[1]{}

\newcommand{\alnote}[1]{}
\newcommand{\amnote}[1]{}
\newcommand{\athotanote}[1]{}
\newcommand{\smnote}[1]{}
\newcommand{\pmnote}[1]{}
\newcommand{\msnote}[1]{}
\newcommand{\mrnote}[1]{}
\newcommand{\aznote}[1]{}
\fi

\lstdefinestyle{myListing}{
  frame=single,
  backgroundcolor=\color{listinggray},
  language=C,
  basicstyle=\ttfamily \footnotesize,
  breakautoindent=true,
  breaklines=true
  tabsize=2,
  captionpos=b,
  aboveskip=0em,
  belowskip=-2em,
}

\lstdefinestyle{myPythonListing}{
  frame=single,
  backgroundcolor=\color{listinggray},
  language=Python,
  basicstyle=\ttfamily \scriptsize,
  breakautoindent=true,
  breaklines=true
  tabsize=2,
  captionpos=b,
}

\ifpdf
\DeclareGraphicsExtensions{.pdf, .jpg, .tif}
\else
\DeclareGraphicsExtensions{.ps,  .eps, .jpg}
\fi

%% file: abstract.tex
The computing systems used by LHC experiments has historically consisted of
the federation of hundreds to thousands of distributed resources, ranging
from small to mid-size resource.  In spite of the impressive scale of the
existing distributed computing solutions, the federation of small to mid-size
resources will be insufficient to meet projected future demands. This paper
is a case study of how the ATLAS experiment has embraced Titan---a DOE
leadership facility in conjunction with traditional distributed high-
throughput computing to reach sustained production scales of approximately
52M core-hours a years. The three main contributions of this paper are:  (i)
a critical evaluation of design and operational considerations  to support
the sustained, scalable and production usage of Titan;  (ii) a preliminary
characterization of a next generation executor for PanDA to support new
workloads and  advanced execution modes; and (iii) early lessons for how
current and future experimental and observational systems can be integrated
with production supercomputers and other platforms in a general and
extensible manner.

%% file: introduction.tex
The Large Hadron Collider (LHC) was created to explore the fundamental
properties of matter. Multiple experiments at LHC have collected and
distributed hundreds of petabytes of data worldwide to hundreds of computer
centers. Thousands of physicists analyze petascale data volumes daily. The
detection of the Higgs Boson in 2013 speaks to the success of the detector
and experiment design, as well as the sophistication of computing systems
devised to analyze the data, which historically, consisted of the federation
of hundreds to thousands of distributed resources, ranging in scale from
small to mid-size resource~\cite{foster2003grid}.

The LHC workloads are comprised of tasks that are independent of each other,
however, the management of the distribution of workloads across many
heterogeneous resources, the effective utilization of resources and efficient
execution of workloads present non-trivial challenges. Many software
solutions have been developed in response to these challenges. The CMS
experiment, devised a solution based around the
HTCondor~\cite{thain2005distributed} software ecosystem. The
ATLAS~\cite{Aad:2008} experiment utilizes the Production and Distributed
Analysis (PanDA) workload management system~\cite{Maeno2011} (WMS) for
distributed data processing and analysis. The CMS and ATLAS experiments
utilize, arguably the largest academic production grade distributed computing
solutions, and have symbolized the paradigm of high-throughput computing
(HTC), i.e., the effective execution of many independent tasks.

In spite of the impressive scale of the ATLAS distributed computing
system---in the number of tasks executed, the number of core hours utilized,
and the number of distributed sites utilized,  demand for computing systems
will soon significantly outstrip current and projected supply.   The data
volumes that will need analyzing in LHC-Run 3 (\(\approx\)2022) and the
high-luminosity era (Run 4) will increase by factors of 10--100 compared to
the current phase (Run 2). There are multiple levels at which this problem
needs to be addressed: the utilization of emerging parallel architectures
(e.g., platforms); algorithmic and advances in analytical methods (e.g., use
of Machine Learning); and the ability to exploit different platforms (e.g.,
clouds and supercomputers).

This paper represents the experience of how the ATLAS experiment has ``broken
free'' of the traditional computational approach of high-throughput computing
on distributed resources to embrace new platforms, in particular
high-performance computers (HPC). Specifically, we discuss the experience of
integrating PanDA WMS with a US DOE leadership machine (Titan) to reach
sustained production scales of approximately 51M core-hours a year.

In doing so, we demonstrate how Titan is more efficiently utilized by the
mixing of small and short-lived tasks in backfill with regular payloads.
Cycles otherwise unusable (or very difficult to use) are used for science,
thus increasing the overall utilization on Titan without loss of overall
quality-of-service. The conventional mix of jobs at OLCF cannot be
effectively backfilled because of size, duration, and scheduling policies.
Our approach is extensible to any HPC with ``capability scheduling''
policies.  We also investigate the use of a pilot-abstraction based task
execution runtime system to flexibly execute ATLAS and other heterogeneous
workloads (molecular dynamics) using regular queues. As such, our approach
provides a general solution and investigation of the convergence of HPC and
HTC execution of workloads.

This work demonstrates a viable production route to delivering large amounts
of computing resources to ATLAS and, in the future, to other experimental and
observational use cases.  This broadens the use of leadership computing while
demonstrating how distributed workflows can be integrated with leadership
resources, and effectively accommodating HTC and HPC workloads
simultaneously.

This paper also provides: (i) a critical evaluation of the many design and
operational considerations that have been taken to support the sustained,
scalable and production usage of Titan for historically high-throughput
workloads, and (ii) early lessons and guidance on designing the next
generation of online analytical platforms~\cite{foap-url},  so that
experimental and observational systems can be integrated with production
supercomputers in a general and extensible manner.

%% file: PandaOverview.tex
PanDA is a Workload Management System (WMS) 
designed to support the execution of distributed workloads and workflows via
pilots~\cite{turilli2017comprehensive}. Pilot-capable WMS enable high
throughput execution of tasks via multi-level scheduling while supporting
interoperability across multiple sites. This is particularly relevant for LHC
experiments, where millions of tasks are executed across multiple sites every
month, analyzing and producing petabytes of data.

The implementation of PanDA WMS consists of several interconnected
subsystems, communicating via dedicated API or HTTP messaging, and
implemented by one or more modules. Databases are used to store stateful
entities like tasks, jobs and input/output data, and to store information
about sites, resources, logs, and accounting.

\begin{figure}
    \centering
    \includegraphics[width=\columnwidth]{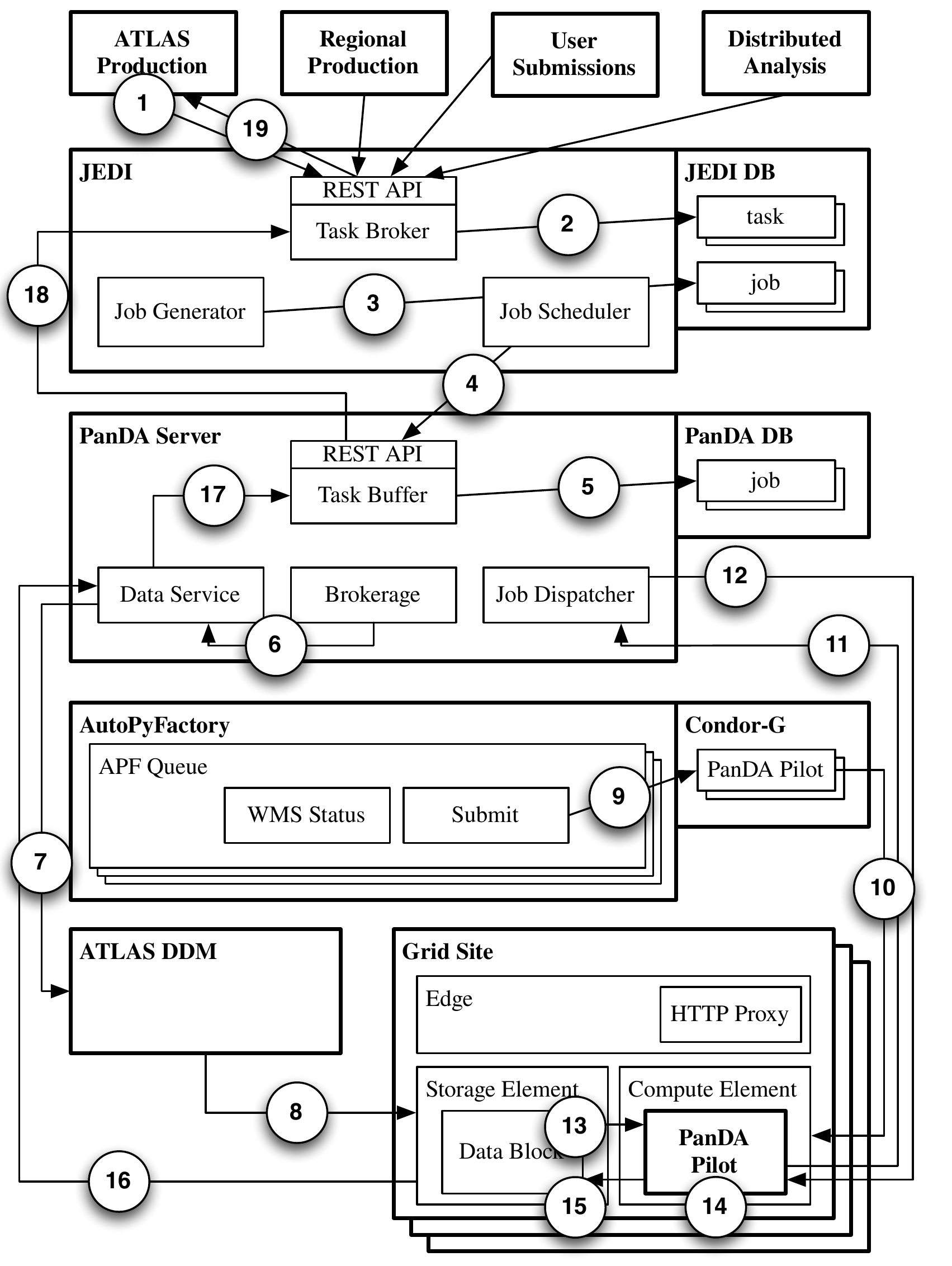}
    \vspace{-0.3in}
    \caption{PanDA WMS architecture. Numbers indicates the JEDI-based
    execution process described in \S\ref{sec:panda_overview}. Several
    subsystems, components, and architectural and communication details are
    abstracted to improve clarity.}\label{fig:architecture}
\end{figure}

Currently, PanDA's architecture has five main subsystems: PanDA
Server~\cite{maeno2011overview},
AutoPyFactory~\cite{caballero2012autopyfactory}, PanDA
Pilot~\cite{nilsson2011atlas}, JEDI~\cite{borodin2015scaling}, and PanDA
Monitoring~\cite{klimentov2011atlas}. Other subsystems are used by some of
ATLAS workflows but we do not discuss them as they are not relevant to an
understanding of how PanDA has been ported to supercomputers. For a full list
of subsystems see Ref.~\cite{panda-wiki_url}. Fig.~\ref{fig:architecture}
shows a diagrammatic representation of PanDA main subsystems, highlighting
the execution process of tasks while omitting monitoring details to improve
readability.

Users submit task descriptions to JEDI (Fig.~\ref{fig:architecture}:1) that
stores them into a queue implemented by a database
(Fig.~\ref{fig:architecture}:2). Tasks are partitioned into jobs of different
size, depending on both static and dynamic information about available
resources (Fig.~\ref{fig:architecture}:3). Jobs are bound to sites with
resources that best match jobs' requirements, and submitted to the PanDA
Server for execution (Fig.~\ref{fig:architecture}:4).

Once submitted to the PanDA Server, jobs are stored by the Task Buffer
component into a global queue implemented as a database
(Fig.~\ref{fig:architecture}:5). When jobs are submitted directly to the
PanDA Server, the Brokerage component is used to bind jobs to available
sites, depending on static information about the resources available for each
site. Jobs submitted by JEDI are already bound to sites so no further
brokerage is needed.

Once jobs are bound to sites, the Brokerage module communicates to the Data
Service module what data sets need to be made available on what site
(Fig.~\ref{fig:architecture}:6). The Data Service communicates these
requirements to the ATLAS DDM (Fig.~\ref{fig:architecture}:7) that, when
needed, replicates data sets on the required sites
(Fig.~\ref{fig:architecture}:8).

Meanwhile, AutoPyFactory defines PanDA Pilots, submitting them to a Condor-G
agent (Fig.~\ref{fig:architecture}:9). Condor-G schedules these pilots
wrapped as jobs to the required sites (Fig.~\ref{fig:architecture}:10).

When a PanDA Pilot becomes available, it requests the Job Dispatcher module
of the PanDA Server for a job to execute (Fig.~\ref{fig:architecture}:11).
The Job Dispatcher interrogates the Task Buffer module for a job that is
bound to the site of that pilot and ready to be executed. Task Buffer checks
the global queue (i.e., the PanDA DB) and, upon availability, returns a job
to the Job Dispatcher. The Job Dispatcher dispatches that job to the PanDA
Pilot (Fig.~\ref{fig:architecture}:12).

Each PanDA Pilot starts a monitoring process on receiving a job and forks a
subprocess to execute the job's payload. Input data are transferred from the
stage-in location (Fig.~\ref{fig:architecture}:13), the job's payload is
executed (Fig.~\ref{fig:architecture}:14) and once completed, output is
transferred to the staging-out location (Fig.~\ref{fig:architecture}:15).

The Data Service module of the PanDA Server tracks and collects the output
generated by each job (Fig.~\ref{fig:architecture}:16), updating jobs'
attributes via the Task Buffer module (Fig.~\ref{fig:architecture}:17). When
the output of all the jobs of a task are retrieved, it is made available to
the user via PanDA Server. When a task is submitted to JEDI, task is instead
marked as done (Fig.~\ref{fig:architecture}:18) and the result of its
execution is made available to the user by JEDI
(Fig.~\ref{fig:architecture}:19).

%% file: PandaEvolution.tex
The upcoming LHC Run 3 will require more resources than the Worldwide LHC
Computing Grid (WLCG) can provide. Currently, PanDA WMS uses more than
600,000 cores at more than 100 Grid sites, with an aggregated performance of
8 petaFLOPS\@. This capacity will be sufficient for the planned analysis and
data processing, but it will be insufficient for the Monte Carlo production
workflow and any extra activity. To alleviate these challenges, ATLAS is
expanding the current computing model to include additional resources such as
the opportunistic use of supercomputers.

PanDA WMS has been designed to support distributed Grid computing. Executing
ATLAS workloads or workflows involves concurrent and/or sequential runs of
possibly large number of jobs, each requiring minimal, if any parallelization
and no runtime communication. Thus, computing infrastructure like WLCG have
been designed to aggregate large amount of computing resources across
multiple sites. While each site may deploy MPI capabilities, usually these
are not used to perform distributed computations.

Currently, ATLAS workloads do not require fast interconnects or specilized
co-processors, but supercomputers tend not to reach 100\% utilization due to
the scheduling of jobs requiring large amount of resources. This offers the
possibility to execute ATLAS-like workloads on supercomputers to increase
utilization and reducing the waste of available resources.

We developed a single-point solution to better understand the problem space
of enabling a WMS designed for HTC to execute production workflows on
resources designed to support HPC\@. The PanDA team developed a job broker to
support the execution of part of the ATLAS production Monte Carlo workflow on
Titan, a leadership-class supercomputer managed by the Oak Ridge Leadership
Computing Facility (OLCF).

\subsection{Architectures, Interfaces and Workloads}\label{ssec:panda-titan}

Titan's architecture, configuration and policies poses several challenges to
the deployment of PanDA\@. The default deployment model of PanDA Pilot is
unfeasible on Titan: PanDA Pilot is required to contact the Job Dispatcher of
the PanDA Server to pull jobs to execute, but this is not possible on Titan
because worker nodes do not offer outbound network connectivity. Further,
Titan does not support PanDA's security model based on certificates and
virtual organizations, making PanDA's approach to identity management
unfeasible. While Titan's data transfer nodes (DTNs) offer wide area network
data transfer, an integration with ATLAS DDM is beyond the functional and
administrative scope of the current prototyping phase. Finally, the specific
characteristics of the execution environment, especially the absence of local
storage on the worker nodes and modules tailored to Compute Node Linux,
require re-engineering of ATLAS application frameworks.

Currently, very few HEP applications can benefit from Titan's GPUs but some
computationally-intensive and non memory-intensive tasks of ATLAS workflows
can be off-loaded from the Grid to Titan. Further, when HEP tasks can be
partitioned into independent jobs, Titan worker nodes can be used to execute
up to 16 concurrent payloads, one per each available core. Given these
constraints and challenges, the  Monte Carlo detector simulation task is most
suitable for execution on Titan at the moment. This type of task is mostly
computational-intensive, requiring less than 2GB of RAM at runtime and small
input data. Detector simulation tasks in ATLAS are performed via
AthenaMP~\cite{aad2010atlas}, the ATLAS software framework integrating the
GEANT4 detector simulation toolkit~\cite{agostinelli2003geant4}. These tasks
account for \(\approx\)60\% of all the jobs on WLCG, making them a primary
candidate for offloading.

Detector simulation is part of the ATLAS production Monte Carlo (MC)
workflow~\cite{rimoldi2006atlas}. The MC workflow consists of four main
stages: event generation, detector simulation, digitization, and
reconstruction. Event generation creates sets of particle four-momenta via
different generators, e.g., PYTHIA, HERWIG, and many others. Geant4 simulates
the ATLAS detector and the interaction between the detector and particles.
Each interaction creates a so-called hit and all hits are collected and
passed on for digitalization, where hits are further processed to mimic the
readout of the detector. Finally, reconstruction operates local pattern
recognition, creating high-level objects like particles and jets.

\subsection{PanDA Broker}\label{ssec:panda_titan}

The lack of wide area network connectivity on Titan's worker nodes is the
most relevant challenge for integrating PanDA WMS and Titan. Without
connectivity, Panda Pilots cannot be scheduled on worker nodes because they
would not be able to communicate with PanDA Server and therefore pull and
execute jobs. This makes impossible to port PanDA Pilot to Titan while
maintaining the defining feature of the pilot abstraction: decoupling
resource acquisition from workload execution via multi-stage scheduling.

The unavailability of pilots is a potential drawback when executing
distributed workloads like MC detector simulation. Pilots are used to
increase the throughput of distributed workloads: while pilots have to wait
in the supercomputer's queue, once scheduled, they can pull and execute jobs
independent from the system's queue. Jobs can be concurrently executed on
every core available to the pilot, and multiple generations of concurrent
executions can be performed until the pilot's walltime is exhausted. This is
particularly relevant for machines like Titan where queue policies privilege
parallel jobs on the base of the number of worker nodes they request: the
higher the number of nodes, the shorter the amount of queue time (modulo
fair-share and allocation policies).

The backfill optimization of Titan's Moab scheduler allows to avoid the
overhead of queue wait times without using pilot
abstraction~\cite{maui_backfill_url}. With this optimization, Moab starts
low-priority jobs when they do not delay higher priority jobs, independent of
whether the low-priority jobs were queued after the high-priority ones.

When the backfill optimization is enabled, users can interrogate Moab about
the number of worker nodes and walltime that would be available to a
low-priority job at that moment in time. If a job is immediately submitted to
Titan with that number of worker nodes and walltime, chances are that Moab
will immediately schedule it, reducing its queue time to a minimum. In this
paper, we call this number of worker nodes and walltime an available
`backfill slot'.

Compared to pilots, backfill has the disadvantage of limiting the amount of
resources that can be requested. Pilots are normal jobs: they can request as
many worker nodes and walltime as a queue can offer. On the contrary, jobs
sized according to an available backfill slot depend on the number of worker
nodes and walltime that cannot be given to any other job at that moment in
time.

At any point in time, the size of an available backfill slot is typically a
small fraction of the total capacity of a resource. Notwithstanding, given
the size of Titan this translates into a substantial capacity. Every year,
about 10\% of Titan's capacity remains unused~\cite{barker2016us},
corresponding to an average of 30,000 unused cores (excluding GPU cores).
This equals to roughly 5\% of the overall capacity of WLCG\@.

\begin{figure}
    \centering
    \includegraphics[width=\columnwidth]{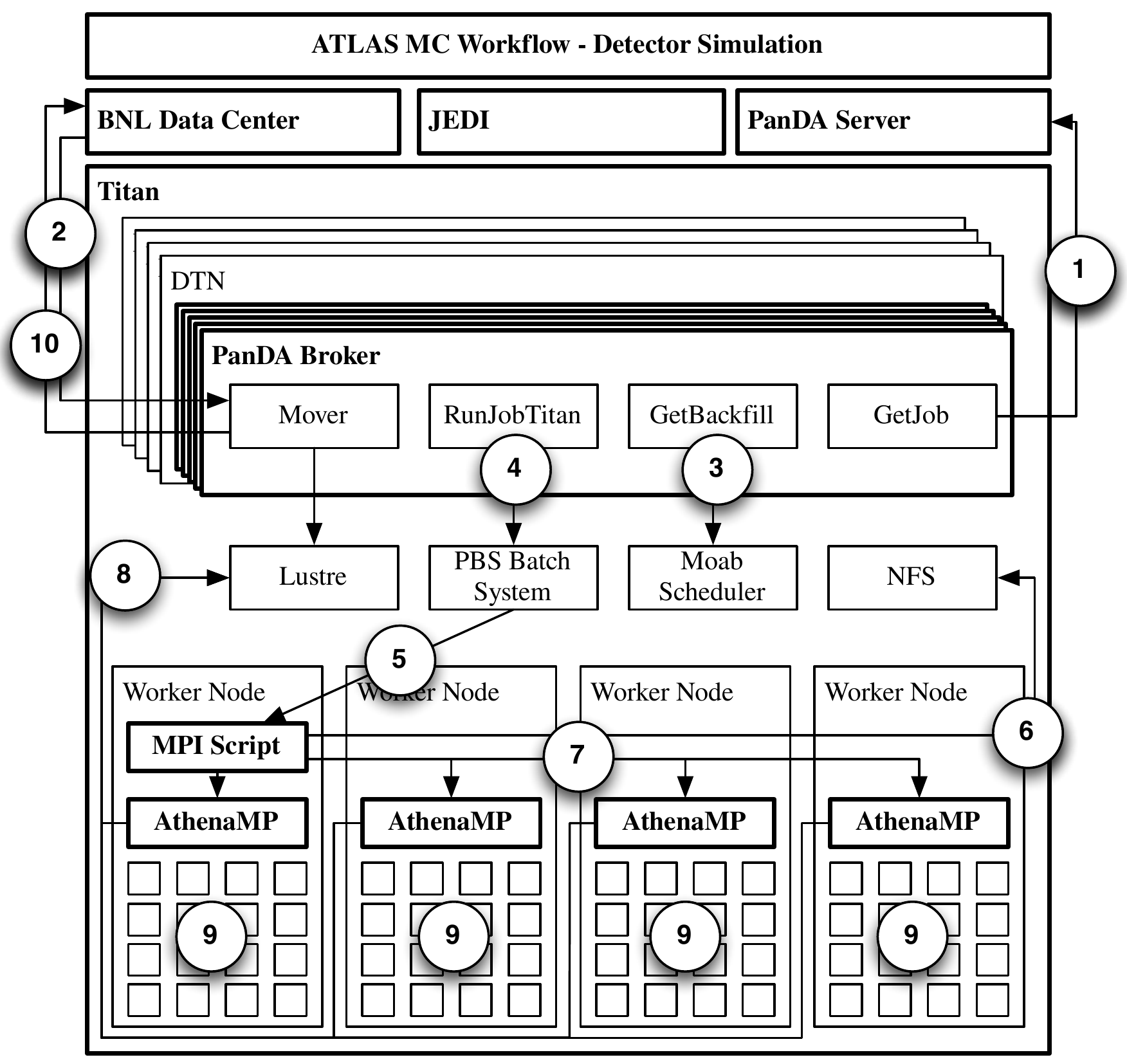}
    \vspace{-0.3in}
    \caption{PanDA Broker architecture as deployed on Titan. Numbers
    indicates the execution process of a detector simulation job described in
    \S\ref{ssec:panda_titan}.}\label{fig:panda_broker}
\end{figure}

Given the communication requirements of PanDA Pilots and the unused capacity
of Titan, PanDA pilot was repurposed to serve as a job broker on the DTN
nodes of Titan (Fig.~\ref{fig:panda_broker}). This prototype called `PanDA
Broker' maintains the core modules of PanDA Pilot and its stand-alone
architecture. This imposes functional trade-offs (e.g., single-threaded
architecture, single MPI PBS script submission) but allows for rapid adoption
and iterative optimization. PanDA Brokers are deployed on DTNs because these
nodes are part of the OLCF infrastructure and can access Titan without RSA
SecureID authentication. DTNs are not part of Titan's worker nodes and,
therefore, are not used to execute Titan's jobs.

Currently, up to 20 PanDA Brokers operate within the existing ATLAS
production software infrastructure, each supporting the execution of MC
detector simulations in 9 steps. Each broker queries the PanDA Server for
ATLAS jobs that have been bound to Titan by JEDI
(Fig.~\ref{fig:panda_broker}:1). Upon receiving jobs descriptions, PanDA
Broker pulls jobs' input files from BNL Data Center to the OLCF Lustre file
system (Fig.~\ref{fig:panda_broker}:2). PanDA Broker queries Titan's Moab
scheduler about the current available backfill slot
(Fig.~\ref{fig:panda_broker}:3) and creates an MPI script, wrapping enough
ATLAS jobs' payload to fit the backfill slot. PanDA Broker submits the MPI
script to the Titan's PBS batch system via RADICAL-SAGA
(Fig.~\ref{fig:panda_broker}:4).

Upon execution on the worker node(s) (Fig.~\ref{fig:panda_broker}:5), the MPI
script initializes and configures the execution environment
(Fig.~\ref{fig:panda_broker}:6), and executes one AthenaMP for each available
work node (Fig.~\ref{fig:panda_broker}:7). AthenaMP retrieves events from
Lustre (Fig.~\ref{fig:panda_broker}:8) and spawns 1 Geant4 event simulation
process on each of the 16 available cores (Fig.~\ref{fig:panda_broker}:9).
Upon completion of each MPI script, PanDA Broker transfer the jobs' output to
BNL (Fig.~\ref{fig:panda_broker}:10), and performs cleanup.

PanDA Broker implementation is resource specific but the ATLAS team has
ported it to other supercomputers, including the HPC2 at the National
Research Center ``Kurchatov Institute''
(NRC-KI)~\cite{belyaev2015integration}, Edison/Cori at the National Energy
Research Scientific Computing Center (NERSC)~\cite{barreiro2016panda}, and
SuperMUC at the Leibniz Supercomputing Centre (LRZ)~\cite{barreiro2016panda}.

\vspace{-0.04in}

%% file: overall.tex
Currently 20 instances of the PanDA Broker are deployed on 4 DTNs, with 5
instances per DTN\@. Each broker submits and manages the execution of 15 to
300 jobs, one job for each Titan worker node, and a theoretical maximum
concurrent use of 96,000 cores. Since November 2015, PanDA Brokers have
operated only in backfill mode, without a defined time allocation, and
running at the lowest priority on Titan. Therefore, ATLAS contributed to an
increase of Titan's utilization.

We evaluate the efficiency, scalability and reliability of the deployment of
PanDA WMS on Titan by characterizing the behavior of both PanDA Broker and
AthenaMP\@. We discuss challenges and limitations of our approach at multiple
levels arising from the specifics of workload, middleware and methods. All
the measurements were performed between January 2016 and February 2017,
hereafter called `experiment time window'.

\subsection{Characterizing the PanDA Broker on
Titan}\label{ssec:broker_titan}

We calculate the total amount of backfill availability over a period of time
by: (i) polling the available backfill slots at regular intervals during that
time window; (ii) converting the number of worker nodes available and their
walltime into core-hours; (iii) summing the number of core-hours. We call
this number of core-hours `total backfill availability'.

\begin{figure}[!t]
    \includegraphics[clip,width=\columnwidth]{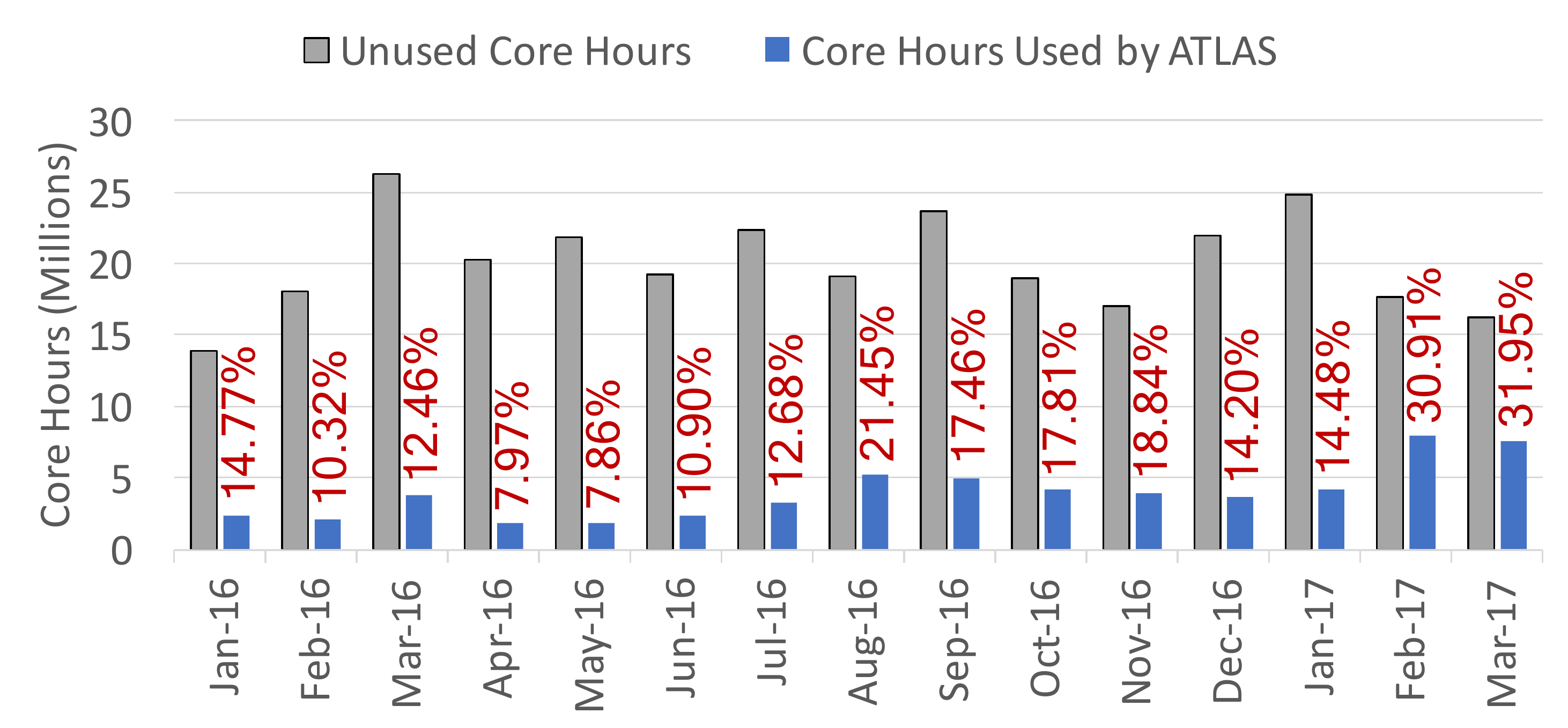}
    \vspace{-0.3in}
    \caption{Titan's total backfill availability: CPU core-hours (gray) and
    CPU core-hours used by ATLAS (blue). GPU core-hours unaccounted for as
    they cannot be used by ATLAS\@. Efficiency of PanDA Brokers defined as
    percentage of total Titan's backfill availability used by ATLAS (Red
    labels).}\label{fig:backfill-utilization}
\end{figure}

Fig.~\ref{fig:backfill-utilization} shows the total backfill availability on
Titan (gray bars) and the number of core-hours of that availability used by
ATLAS (blue bars) during the experiment time window. ATLAS consumed a total
of 51.4M core-hours, for an average of \(\approx\)3.7M core-hours a month.

PanDA Brokers' efficiency (Fig.~\ref{fig:backfill-utilization}, red labels)
is defined as the fraction (or percentage) of core-hours utilized by the
PanDA Brokers of Titan’s total backfill availability during the experiment
time window. The  average efficiency was 18\%, with a minimum efficiency of
7.8\% (May 2016) and a maximum efficiency of 30.9\% (Feb. 2017, excluding the
preliminary results of March). The total backfill availability was
\(\approx\)21.5M in April 2016, and 17.6M in February 2017. This suggests
that the efficiency is invariant of total backfill availability.

During the experiment time window, about 2.25M detector simulation jobs were
completed on Titan, for a total of 225M events processed. This is equivalent
to 0.9\% of all the 250M detector simulations performed by ATLAS in the same
period of time, and 3.5\% of the 6.6B events processed by those jobs. These
figures confirms the relevance of supercomputers' resource contribution to
the LHC Run 2, especially when accounting for the amount of unused total
backfill availability and the improvement of PanDA efficiency across the
experiment time window.

On February 2017, PanDA Brokers used almost twice as much total backfill
availability than in any other month (preliminary results for March 2017
displayed in Fig.~\ref{fig:backfill-utilization} confirm this trend). No
relevant code update was made during that period and logs indicated that the
brokers were able to perform faster. This is likely due to hardware upgrades
on the DTNs. The absence of continuous monitoring of those nodes does not
allow to quantify bottlenecks but spot measurements of their load indicate
that a faster CPU and better networking were likely responsible for the
improved performance. Investigations showed an average CPU load of 3.6\% on
the upgraded DTNs, as opposed to the ``high'' utilization reported by OLCF
for the previous DTNs. As such, further hardware upgrades seem unlikely to
improve significantly the performance of PanDA Brokers.\mtnote{Possibly, add
a small paragraph summarizing why at the moment we are consuming up to 30\%
of the available backfill.}

Every detector simulation executed on Titan process 100 events. This number
of events is consistent with the physics of the use case and with the average
duration of backfill availability. The duration of a detector simulation is a
function of the number of events simulated but not all events take the same
time to be simulated. One event simulation takes from \(\approx\)2 to
\(\approx\)40 minutes, with a mean of \(\approx\)14 minutes. Considering that
each worker node process up to 16 events concurrently, 100 events takes an
average of 105 minutes to process. As such, PanDA brokers do not use backfill
availability with less than 105 minutes walltime.

\begin{figure}[!t]
    \includegraphics[clip,width=\columnwidth]{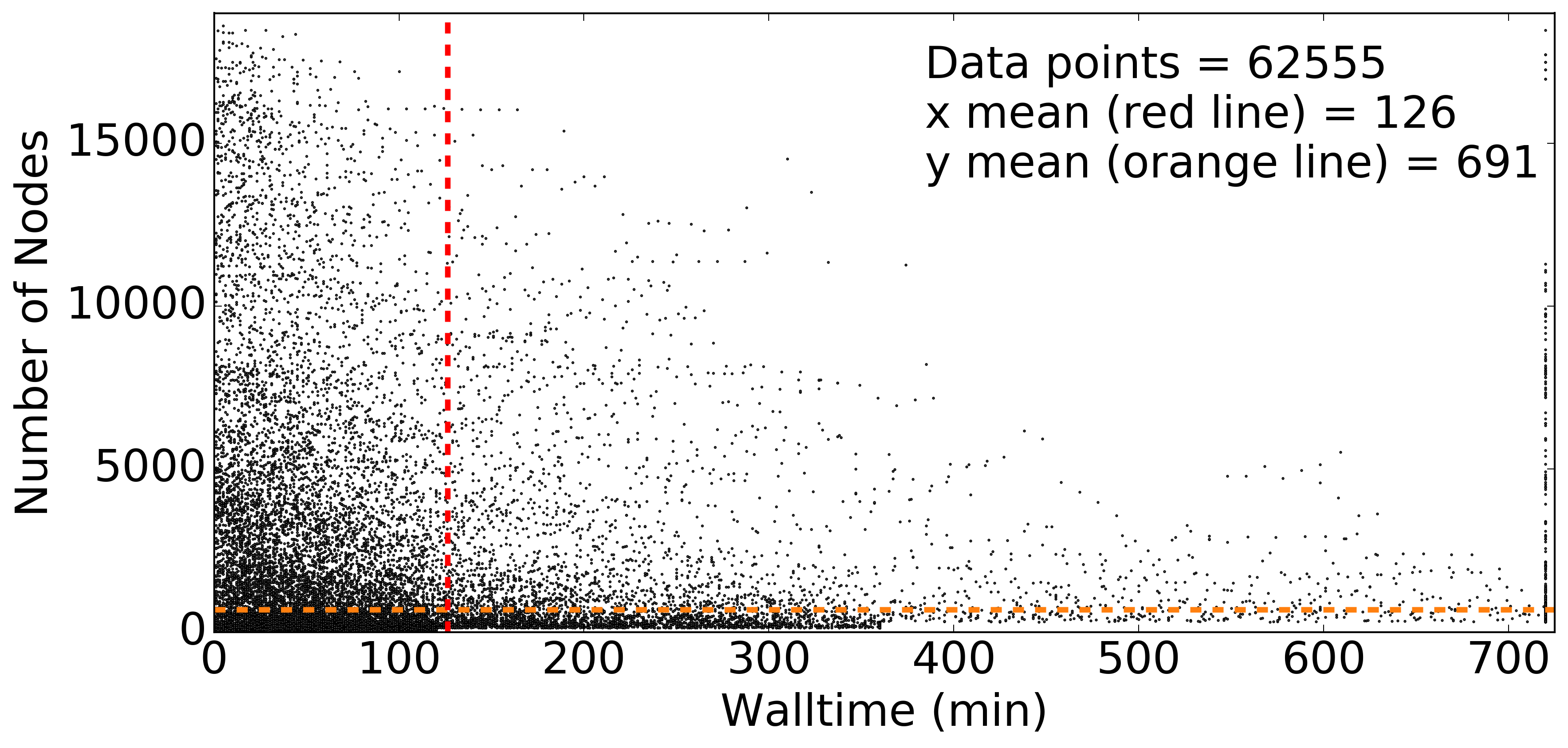}
    \vspace{-0.3in}
    \caption{62555 measures of Backfill availability on Titan during the
    experiment time window. Mean number of work nodes available 691; mean
    walltime available 126 minutes.}\label{fig:backfill-distrib}
\end{figure}

Fig.~\ref{fig:backfill-distrib} shows backfill availability on Titan as a
function of number of nodes and the time of their availability (i.e.,
walltime). We recorded these data by polling Titan's Moab scheduler at
regular intervals during the experiment window time. The mean number of nodes
was 691, and their mean walltime was 126 minutes. Detector simulations of 100
events, enable to use down to 5/6 of the mean walltime of backfill
availability. As such, it offers a good compromise for PanDA Broker
efficiency.

PanDA Broker could fit the number of events to the walltime of each available
backfill slot on the base of the distributions of the time taken by one event
to be simulated. That specific number of event could then be pulled from the
PanDA Event service~\cite{calafiura2015atlas} and given as input to one or
more simulations. Once packaged into the MPI script submitted to titan's PBS
batch system, these simulations would better fit their available backfill
slot, contributing to increase the efficiency of PanDA Brokers.

The transition from a homogeneous to a heterogeneous number of events per
detector simulation has implications for the application layer. An even
number of events across simulations makes it easier to partition, track and
package events across simulations, especially when they are performed on both
the Grid and Titan. A homogeneous number of events also helps to keep the
size and duration of other stages of the MC workflow
(\S\ref{ssec:panda-titan}) more uniform. Further analysis is needed to
evaluate the trade offs between increased efficiency of resource utilization
and the complexity that would be introduced at the application layer.

Currently, each PanDA Broker creates, submits, and monitors a single MPI PBS
script at a time. This design is inherited from PanDA Pilot where a single
process is spawn at a time to execute the payload. As a consequence, the
utilization of a larger portion of Titan's total backfill availability
depends on the the number of concurrent PanDA Brokers instantiated on the
DTNs: When all the 20 PanDA Brokers have submitted a MPI PBS script, further
backfill availability cannot be used.

In August 2016, increasing the number of concurrent PanDA brokers from 4 to
20 markedly improved efficiency (see Fig.~\ref{fig:backfill-utilization}) but
further research is ongoing to understand whether an even greater number of
brokers would yield even greater efficiency. This research focuses on
evaluating the overheads of input/output files staging, including its impact
on DTNs, and on an alternative design of PanDA Broker that enables the
concurrent submission of multiple MPI scripts~\cite{barreiro2016panda}. The
understanding will contribute to improving the efficiency of PanDA Brokers
beyond the 30\% limit showed in Fig.~\ref{fig:backfill-utilization}.

The current design and architecture of the PanDA Broker is proving to be as
reliable as PanDA Pilot when used on the WLCG\@. Between Jan 2016 and Feb
2017, the overall failure rate of all the ATLAS detector simulation jobs was
14\%, while the failure rate of jobs submitted to Titan was a comparable
13.6\%. PanDA Brokers were responsible for around the 19\% of the failures,
compared to the 29\% of failures produced by the JobDispatcher module of the
PanDA Server, and the 13\% failures produced by the Geant4 toolkit. 

\subsection{Characterizing the Detector Simulation on
Titan}\label{ssec:athenamp_titan}

We use two main parameters to measure the performance of the detector
simulation jobs submitted to Titan: (i) the time taken to setup AthenaMP\@;
and (ii) the distribution of the time taken by Geant4 to simulate a certain
number of events.

AthenaMP has an initialization and configuration stage. At initialization
time, AthenaMP is assembled from a very large set of shared libraries,
depending on the type of payload that will have to be computed. Once
initialized, every algorithm and service of AthenaMP is configured via Python
scripts. Both these operations result in read operations on the filesystem
shared between the worker nodes and the DTNs, including the operations
required to access small python scripts.

Initially, all the shared libraries and the python scripts of AthenaMP were
stored on the OLCF Spider 2 Lustre file system. However, the I/O patterns of
the initialization and configuration stages degraded the performance of the
filesystem. This was addressed by moving the AthenaMP distribution to a
read-only NFS directory, shared among Titan's DTNs and worker nodes. NFS
eliminated the problem of metadata contention, improving metadata read
performance from a worse case scenario of 6,300s on Lustre to \(\approx\)225s
on NFS\@. While these figures are specific to OLCF and Titan, reengineering
of AthenaMP could improve startup time on every platform.

Once initialized and configured, AthenaMP is used to execute 16 concurrent
Geant4 simulators on a Titan's worker node. Geant4 requires to read events
descriptions from a filesystem and simulate them as they would happen within
the ATLAS detector. We characterized both the compute performance of the
simulation and the impact of acquiring event descriptions on the filesystem.

\begin{figure}[htp]
    \includegraphics[clip,width=\columnwidth]{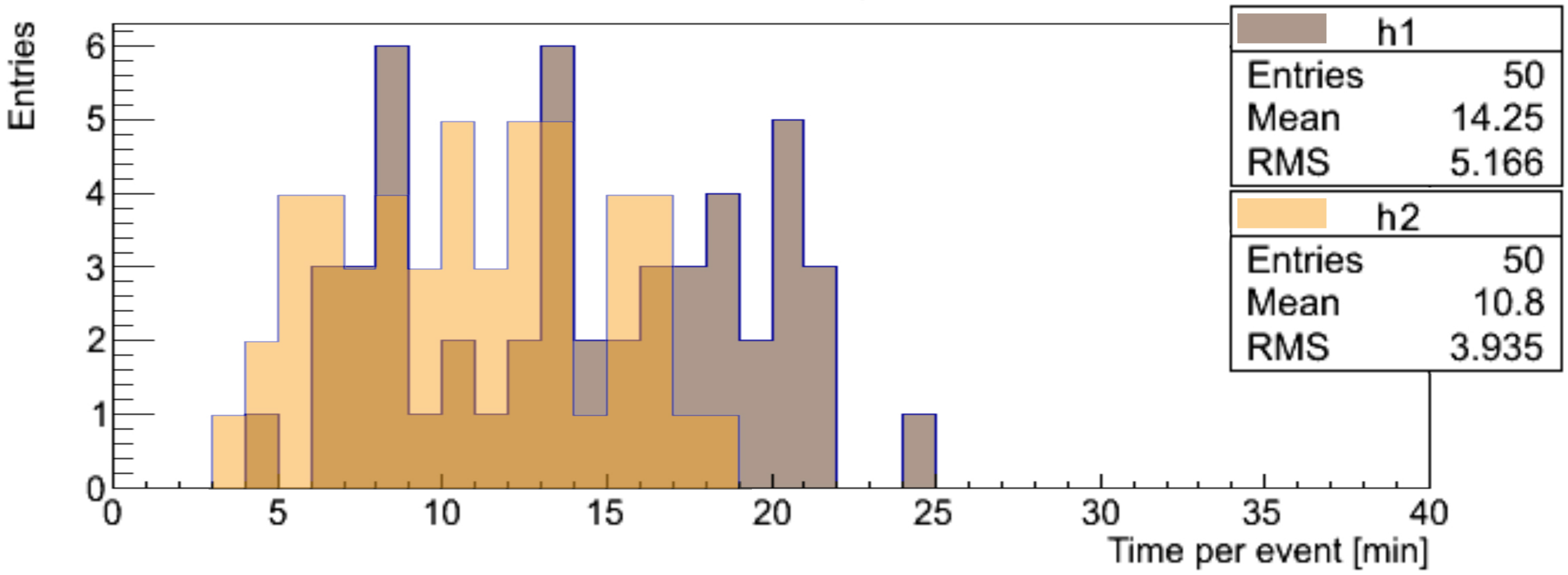}
    \vspace{-0.3in}
    \caption{Distributions of the time taken to simulate one event when
    placing 2 simulations (h1) or 1 simulation (h2) per Titan's CPU\@. 2
    simulation use 16 cores per node, 1 simulation 8. 50 Events; 1 Titan
    worker nodes; 16 work threads per node; 100 events per
    node.}\label{fig:comparison-8-16cores}
\end{figure}

The AMD Opteron 6274 CPU used on Titan has 16 cores, divided into 8 compute
units. Each compute units has 1 floating point (FP) scheduler shared between
2 cores. When using 16 cores for FP-intensive calculations, each pair of
cores competes for a single FP scheduler. This creates the overhead shown in
Fig.~\ref{fig:comparison-8-16cores}: the mean runtime per event for 8
concurrent simulations computing 50 events is 10.8 minutes, while for 16
simulations is 14.25 minutes (consistent with the measured distribution of
the duration of event simulation). Despite an inefficiency of almost 30\%,
Titan's allocation policy based on number of worker nodes used instead of
number of cores does not justify the use of 1/2 of the cores available.

A performance analysis of Titan's AMD CPUs for detector simulations also
helps to compare Titan and Grid site performance. Usually, Grid sites expose
resources with heterogeneous CPU architectures and 8 (virtual) cores, while
Titan's offer an homogeneous 16 cores architecture. We used the rate of
events processes per minute as a measure of the efficiency of executing the
same detector simulation on Titan or Grid sites. Comparisons of the
efficiencies of Titan to the BNL and SIGNET Grid sites, normalized for 8
cores, show that the effective performance per-core at Titan is
\(\approx\)0.57 event per minute, roughly 1/2 of BNL and  1/3 of SIGNET\@.

The differences in performance between Titan and the BNL and SIGNET Grid
sites are due to the FP scheduler competition but mainly to the availability
of newer processors. The heterogeneity of the Grid sites' CPUs explain the
higher performance variance we observed compared to the performance
consistency we measured on Titan. This difference in performance is
compensated by the amount of resources available on Titan (capable of
executing up to 30\%/year of the ATLAS detector simulations) and by the lack
of further resources available to WLCG\@. Also, it should be noted that Titan
is at the end of its life-cycle and that Summit, Titan's successor, will
offer cutting-edge performances.

We studied the impact of acquiring event descriptions on Lustre by analyzing
1,175 jobs ran on the week of 10/25/2016, for a total of 174 hours.
Table~\ref{panda-olcf-stats} shows the overall statistical breakdown of the
observed file I/O. ATLAS used between 1 and 300 worker nodes, and 35 on
average. 75\% of the jobs run by ATLAS consumed less than 25 nodes and 92\%
less than 100. During the 174 hours of data collection, 6.75 ATLAS jobs were
executed on average per hour, each job running for an average of 1.74 hours.
Every job read less than 250 GB and wrote less than 75 GB of data and, on
average, each job read 20 GB and wrote 6 GB of data.

\begin{table*}[t]
\centering
\begin{tabular}{lllllllll} & Num. Nodes & Duration (s) & Read (GB) & Written
 (GB) & GB Read/nodes & GB Written/nodes & \(open()\) & \(close()\) \\ Min &
 1 & 1,932 & 0.01 & 0.03 & 0.00037 & 0.02485 & 1,368 & 349 \\ Max & 300 &
 7,452 & 241.06 & 71.71 & 0.81670 & 0.23903 & 1,260,185 & 294,908 \\ Average
 & 35.66 & 6,280.82 & 20.36 & 6.87 & 0.38354 & 0.16794 & 146,459.37 &
 34,155.74 \\ Std. Dev. & 55.33 & 520.99 & 43.90 & 12.33 & 0.19379 & 0.03376
 & 231,346.55 & 53,799.08
\end{tabular}
\caption{The Statistical breakdown of the I/O impact of 1,175 jobs ATLAS
executed at OLCF for the week of 10/25/16}\label{panda-olcf-stats}
\end{table*}

ATLAS jobs are read heavy: On average, the amount of data read per worker
node is less than 400 MB, while the amount of data written is less than 170
MB\@. Distributions of read and written data are different: The read
operation distribution per job shows a long tail, ranging from 12.5 GB to 250
GB, while the written amount of data has a very narrow distribution.

The metadata I/O breakdown shows that ATLAS jobs yield 23 file \(open()\)
operations per second (not including file \(stat()\) operations) and 5 file
\(close()\) operations per second, with similar distributions. On average,
the maximum number of file \(open()\) operations per job is \(\approx\)170/s
and the maximum number of file \(close()\) operations is \(\approx\)39/s. For
the 1,175 ATLAS jobs observed, the total number of file \(open()\) operations
is 172,089,760 and the total number of file \(close()\) operations is
40,132,992. The difference between these two values is under investigation: a
possible explanation is that ATLAS jobs don't call a file
\(close()\) operation for every file \(open()\) issued.

Overall, the average time taken to read events from input files stored on
Lustre is 1,320, comparable to the time taken to read the file required by
assembling AthenaMP from NFS\@. Preliminary investigation shows that this
time could be reduced to 40 seconds by loading the event descriptions into
the RAM disk available on each worker node. Events descriptions could be
transferred from Lustre to the RAM disk while configuring and initializing
AthenaMP, almost halving the time currently required by initiating a Geant4
simulation.

%% file: nge.tex
As explained in \S\ref{sec:panda_titan}, PanDA Broker was designed to
maximize code reutilization of PanDA Pilot. This allowed for rapid adoption
and incremental optimization while enabling the parallel development of a
more general solution for executing ATALS workloads on HPC resources.

The lack of pilot capabilities in PanDA Broker impacts both the efficiency
and the flexibility of PanDA's execution process. Pilots could improve
efficiency by increasing throughput and enabling greater backfill
utilization. Further, pilots would make it easier to support heterogeneous
workloads.

The absence of pilots makes the scheduling of multiple generations of
workload on the same PBS job impossible: once a statically defined number of
detector simulations are packaged into a PBS job and this job is queued on
Titan, no further simulations can be added to that job. New simulations have
to be packaged into a new PBS job that needs to be submitted to Titan based
upon backfill and PanDA Brokers availability.

The support of multiple generations of workload would enable more efficient
use of the backfill availability walltime. Currently, when a set of
simulations ends, the PBS job also ends, independent of whether more
wall-time would still be available. With a pilot, additional simulations
could be executed to utilize all the available wall-time, while avoiding
further job packaging and submission overheads.

Multiple generations would also relax two assumptions of the current
execution model: knowing the number of simulations before submitting the MPI
script, and having a fixed number of events per simulation (currently 100).
Pilots would enable the scheduling of simulations independently from whether
they were available at the moment of submitting the pilot. Further,
simulations with a varying number of events could be scheduled on a pilot,
depending on the amount of remaining walltime and the distribution of
execution time per event, as shown in \S\ref{ssec:panda_titan},
Fig.~\ref{fig:comparison-8-16cores}. These capabilities would increase the
efficiency of the PanDA Broker when there is a large difference between the
number of cores and walltime.

Pilots can offer a payload-independent scheduling interface while hiding the
mechanics of coordination and communication among multiple worker nodes. This
could eliminate the need for packaging payload into MPI scripts within the
broker, greatly simplifying the submission process. This simplification would
also enable the submission of different types of payload, without having to
develop a specific PBS script for each payload. The submission process would
also be MPI-independent, as MPI is used for coordination among multiple
worker nodes, not by the payload.

\subsection{Implementation}\label{sec:arch}

The implementation of pilot capabilities within the PanDA Broker require
quantification of the effective benefits that it could yield and, on the base
of this analysis, a dedicated engineering effort. We developed a prototype of
a pilot system capable of executing on Titan to study experimentally the
quantitative and qualitative benefits that it could bring to PanDA\@. We
called this prototype Next Generation Executor (NGE).

\begin{figure}
  \centering
	\includegraphics[width=\columnwidth]{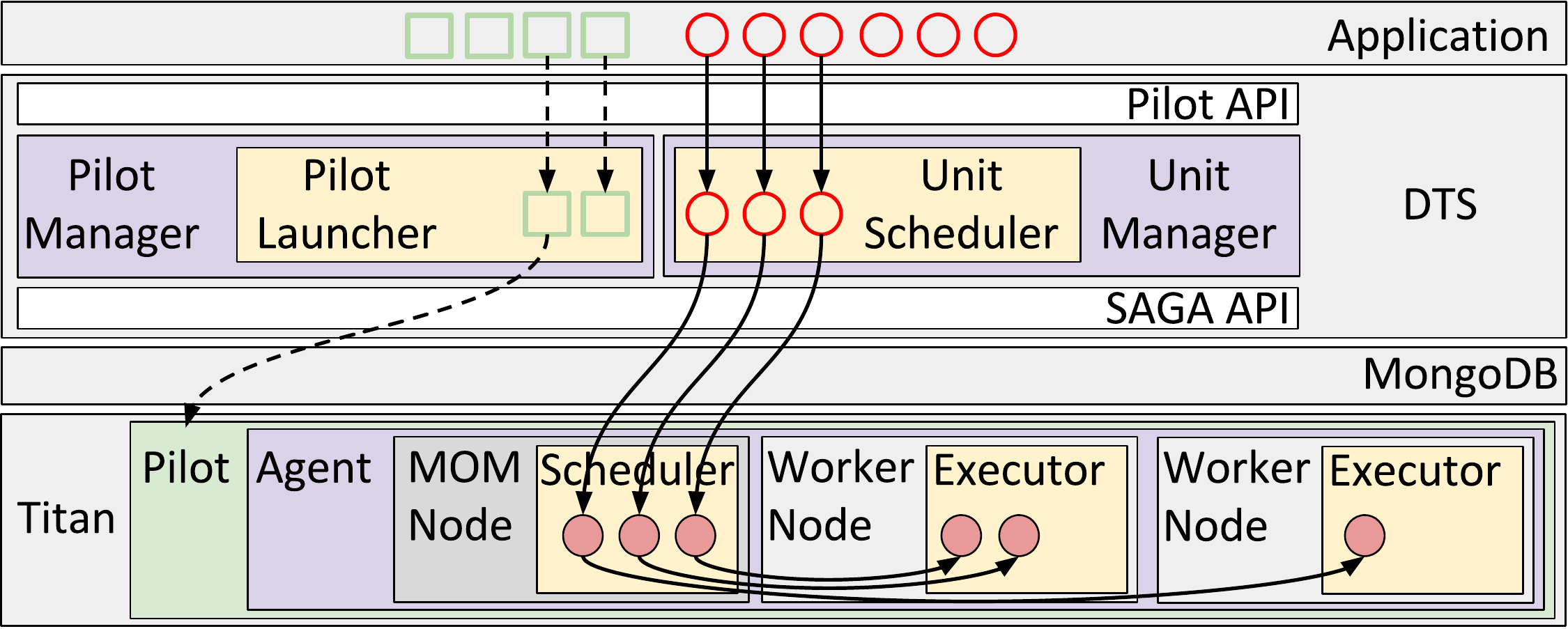}
	\vspace{-0.3in}
	\caption{NGE Architecture: The PilotManager and UnitManager reside on a
  	DTN while the Pilot Agent is executed on a worker node. Color coding:
  	gray for entities external to NGE\@; white for APIs; purple for NGE's
  	modules; green for pilots; yellow for module's components.}
\label{fig:arch-overview}
\end{figure}

NGE is a runtime system to execute heterogeneous and dynamically determined
tasks that constitute workloads. Fig.~\ref{fig:arch-overview} illustrates its
current architecture as deployed on Titan: the two management modules (Pilot
and Unit) represent a simplified version of the PanDA Broker while the agent
module is the pilot submitted to Titan and executed on its worker nodes. The
communication between PanDA Broker and Server is abstracted away as it is not
immediately useful to evaluate the performance and capabilities of a pilot on
Titan.

NGE exposes an API to describe workloads (Fig.~\ref{fig:arch-overview}, green
squares) and pilots (Fig.~\ref{fig:arch-overview}, red circles), and to
instantiate a PilotManager and a UnitManager. The PilotManager submits pilots
to Titan's PBS batch system via SAGA API (Fig.~\ref{fig:arch-overview}, dash
arrow). Once scheduled, the Pilot Agent is bootstrapped on a MOM node and the
Agent's Executors on worker nodes.

The UnitManager and the Pilot Agent communicate via a database instantiated
on a DTN so as to be reachable by both modules. The UnitManager schedules
units to the Agent's Scheduler (Fig.~\ref{fig:arch-overview}, solid arrow)
and the Agent's Scheduler schedules the units on one or more Agent's
Executor.

The Pilot Agent uses the Open Run-Time Environment (ORTE) for communication
and coordination of the execution of units. This environment is a critical
component of the OpenMPI implementation~\cite{castain05:_open_rte}. ORTE is
able to minimize the system overhead while submitting tasks by avoiding
filesystem bottlenecks and race conditions with network sockets.

%% file: Experiments.tex
\subsection{Experiments}\label{sec:ngeExp}

We designed experiments to characterize the performance of the NGE on Titan,
with an emphasis on understanding its overhead and thus the cost of
introducing new functionalities. We perform three groups of experiments in
which we investigate the weak scalability, weak scalability with multiple
generation, and strong scalability of the NGE\@.

Each experiment entails executing multiple instances of AthenaMP to simulate
a pre-determined number of events. All the experiments have been performed by
configuring AthenaMP to use all the 16 cores of Titan's worker nodes.

We measured the execution time of the pilots and of AthenaMP within them,
collecting timestamps at all stages of the execution. Experiments were
performed by submitting pilots to Titan's batch queue. The turnaround time of
an individual run is determined by queue waiting times. Since we are
interested only in the performances of the NGE, we removed queue time from
our statistics.

\subsubsection{Weak scalability}

In this experiment we run as many AthenaMP instances (hereafter referred to
as tasks) as the number of nodes controlled by the pilot. Each AthenaMP
simulates 100 events, requiring \(\sim 4200\) seconds on average.

Tasks do not wait within the Agent's queue since one node is available to
each AthenaMP instance. Task execution  overheads result primarily from three
factors: (i) the initial bootstrapping of the pilot on the nodes; (ii) the
UnitManager's dispatching of units (tasks) to the agent; and (iii) time for
the Agent to bootstrap all the tasks on the nodes.

\begin{figure}[!t]
    \includegraphics[height=4.5cm,width=\columnwidth]{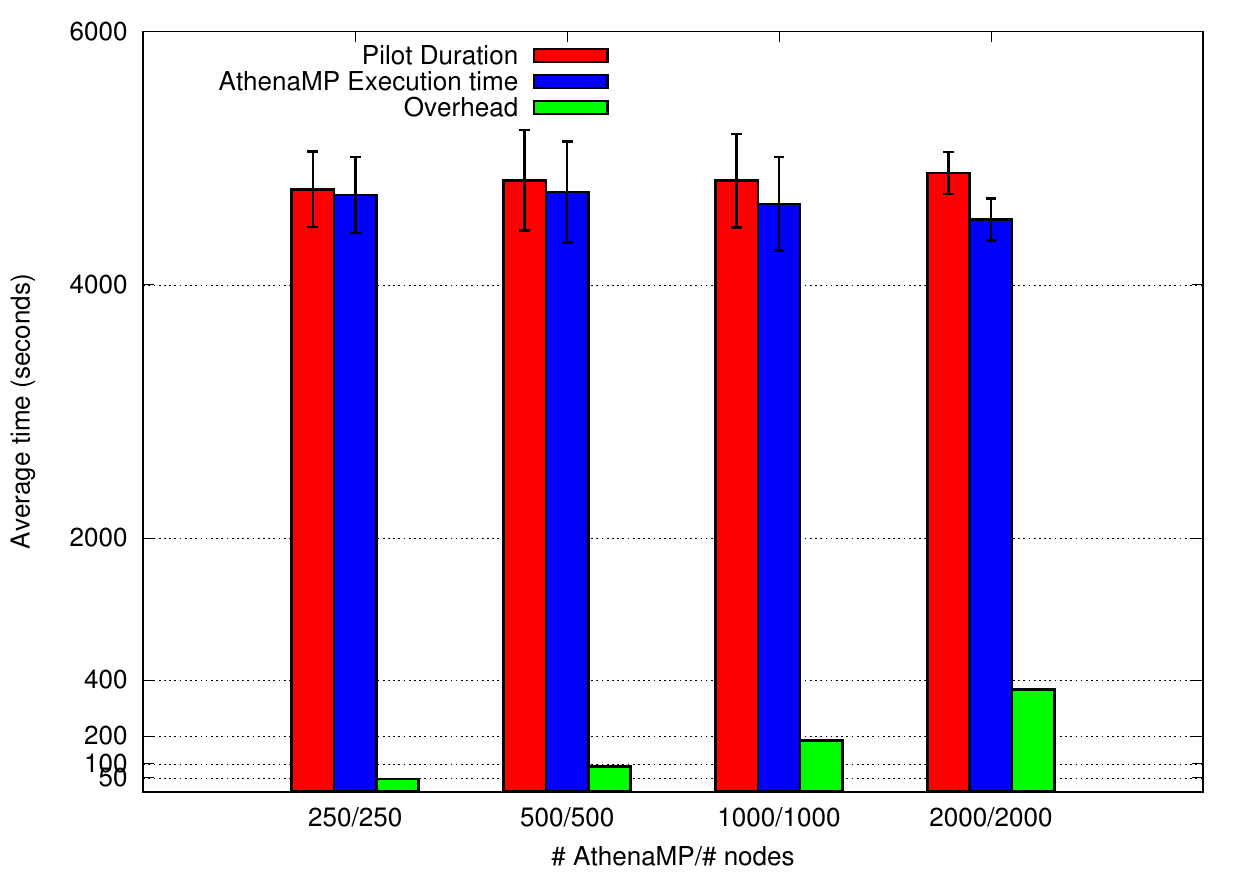}
   	\vspace{-0.3in}
    \caption{Weak scalability: average pilot duration, average duration of
    one AthenaMP execution, and pilot's overhead as a function of pilot sizes
    (200, 500, 1000 and 2000 nodes).}\label{fig:weakScal1a}
\end{figure}

We tested pilots with 250, 500, 1000 and 2000 worker nodes and 2 hours
walltime. The time duration is determined by the Titan's walltime policy.
Fig.~\ref{fig:weakScal1a} depicts the average pilot duration, the average
execution time of AthenaMP, and the pilot overhead as function of the pilot
size.

We observe that, despite some fluctuations due to external factors (e.g.,
Titan's shared filesystem and the shared database used by the NGE), the
average execution time of AthenaMP ranges between 4500 and 4800 seconds. We
also observe that in all the cases the gap between AthenaMP execution times
and the pilot durations is minimal, although it slightly increases with the
pilot size. We notice that NGE's overhead grows linearly with the number
of units.

\subsubsection{Weak scalability with multiple generation}

The NGE provides an important new capability of submitting multiple
generations of tasks to the same pilot. In order to investigate the cost
of doing so, we performed a variant of the weak scalability experiments. This
stresses the pilot's components, as new tasks are scheduled for execution on
the Agent while other tasks are still running.

In these experiments, we run five AthenaMP instances per node. As these
experiments are designed to investigate the overhead of  scheduling and
bootstraping of AthenaMP instances, the number of events simulated by each
AthenaMP task was reduced to sixteen such that the running time of each
AthenaMP was \(\approx\)1,200 seconds on average. This experiment design
choice does not affect the objectives or accuracy of the experiments, but
allows us to scale experiments to large node counts by conserving project
allocation.

\begin{figure}[!t]
    \includegraphics[height=4.5cm,width=\columnwidth]{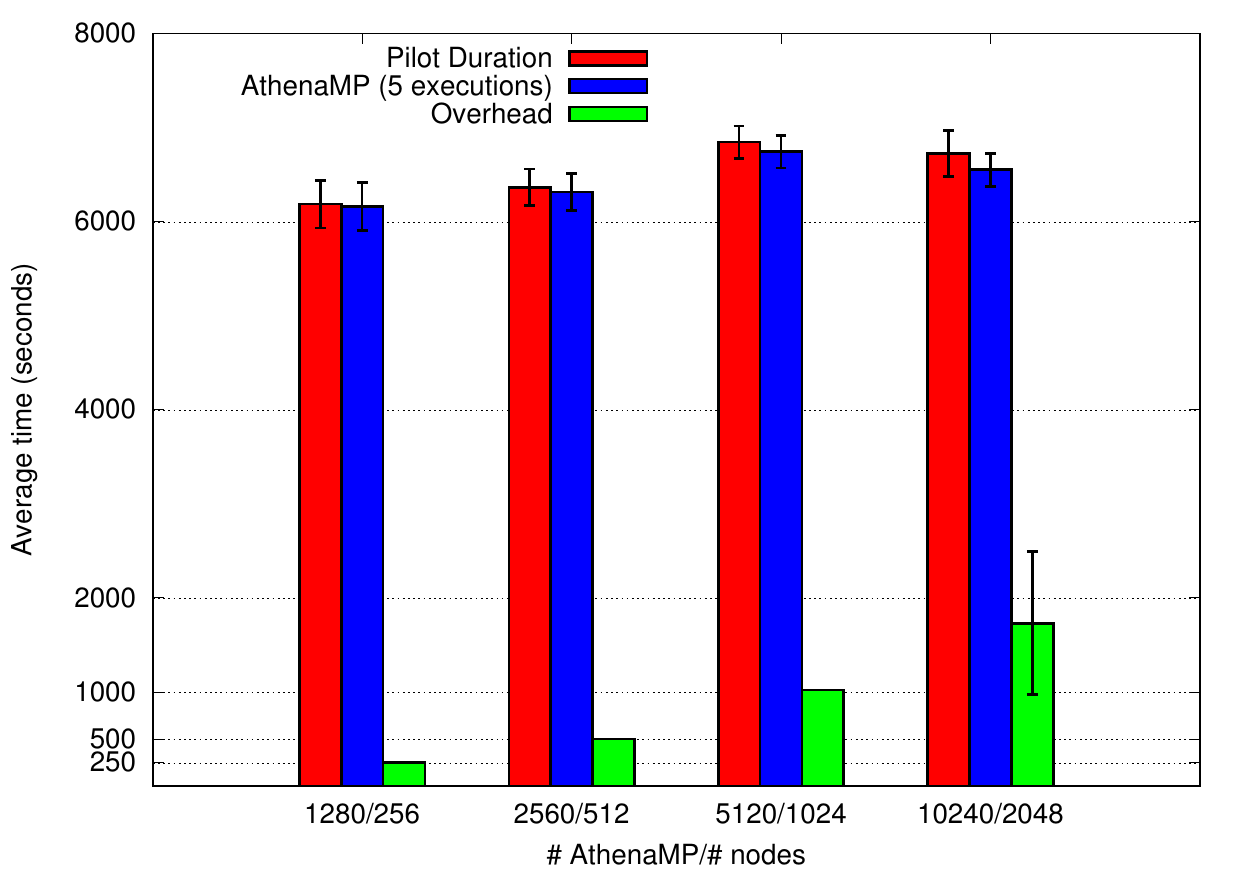}
    \vspace{-0.3in}
    \caption{Weak scalability with multiple generations (where each
    generation has approximately 1/6th the number of events compared to Fig.
    7): average pilot duration, average duration of sequential AthenaMP
    executions, and pilot's overhead for pilot with 256, 512, 1024 and 2048
    nodes.}\label{fig:weakScal2a}
\end{figure}

We ran pilots with 256, 512, 1024 and 2048 worker nodes and 3 hours walltime.
Fig.~\ref{fig:weakScal2a} depicts the average pilot duration, the average
execution time of five generations of AthenaMP, and the corresponding
overhead. The difference between the two durations is more marked than in the
previous experiments. Despite this, we notice that the growth of the overhead
is consistent with the increment of the number of tasks per node for pilots
with 256, 512 and 1024 worker nodes, and less than linear for the pilot with
2048 worker nodes.

\subsubsection{Strong scalability}

We investigate strong scalability by running the same number of tasks for
different pilot sizes. We used 2048 AthenaMP instances and pilots with 256,
512, 1024 and 2048 nodes. Thus, the number of AthenaMP instances per node
(i.e., generations) is eight for the smallest pilot (256 nodes), one for the
largest pilot (2048 nodes), and with the number of generations of AthenaMP
decreasing with the pilot size. These experiments are designed to investigate
whether pilot overhead is affected by the degree of concurrency within the
pilot and/or the number of tasks. Each AthenaMP instance simulates sixteen
events as in the previous experiment.

\begin{figure}[!t]
    \includegraphics[height=4.5cm,width=\columnwidth]{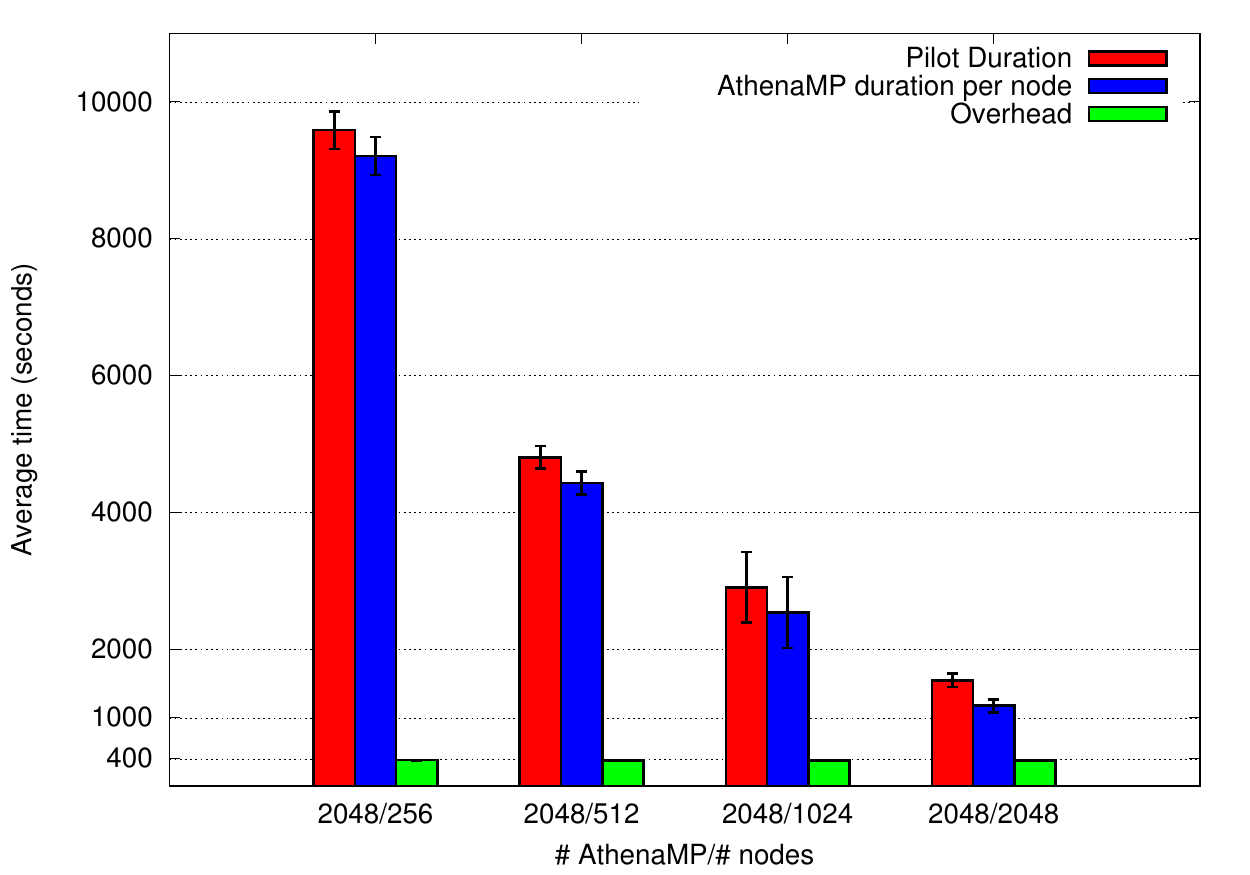}
    \vspace{-0.3in}
    \caption{Strong scalability: Average pilot duration, average duration of
    sequential AthenaMP executions, and pilot's overhead for pilots with 256,
    512, 1024 and 2048 nodes.}\label{fig:strongScala}
\end{figure}

Fig.~\ref{fig:strongScala} shows the average pilot duration and the average
execution time of possibly sequential AthenaMP instances. We notice that the
difference between the pilot duration and the AthenaMP execution times is
almost constant for all the pilot sizes, although the overall duration of the
pilot decreases linearly with the pilot size.

\vspace{-0.05in}

%% file: bckground.tex
Several pilot-enabled WMS were developed for the LHC experiments:
AliEn~\cite{Bagnasco2010} for ALICE\@; DIRAC~\cite{Paterson2010} for LHCb;
GlideinWMS~\cite{sfiligoi2008glideinwms} for CMS\@; and
PanDA~\cite{maeno2014evolution} for ATLAS\@. These systems implement similar
design and architectural principles: centralization of task and resource
management, and of monitoring and accounting; distribution of task execution
across multiple sites; unification of the application interface; hiding of
resource heterogeneity; and collection of static and sometimes dynamic
information about resources.

AliEn, DIRAC, GlideinWMS and PanDA all share a similar design with two types
of components: the management ones facing the application layer and
centralizing the capabilities required to acquire tasks' descriptions and
matching them to resource capabilities; and resource components used to
acquire compute and data resources and information about their capabilities.
Architecturally, the management components include one or more queue and a
scheduler that coordinates with the resource modules via push/pull protocols.
All resource components include middleware-specific APIs to request for
resources, and a pilot capable of pulling tasks from the management modules
and executing them on its resources.

AliEn, DIRAC, GlideinWMS and PanDA also have similar implementations. These
WMS were initially implemented to use Grid resources using the Condor
software ecosystem~\cite{thain2005distributed}. Accordingly, all LHC WMS
implemented Grid-like authentication and authorization systems and adopted a
computational model based on distributing a large amount of single/few-cores
tasks across hundreds of sites.

All LHC experiments produce and process large amounts  of data from actual
collisions in the accelerator and from their simulations. Dedicated,
multi-tiered data systems have been built to store, replicate, and
distributed these data. All LHC WMS interface with these systems to move data
to the sites where related compute tasks are executed or to schedule compute
tasks where (large amount of) data are already stored.

It is interesting to note that most WMS developed to support LHC experiments
are gradually evolving towards integrating HPC resources, though none have
reached sustained operational usage at the scales that PanDA has achieved for
ATLAS on Titan.

%% file: conclusion.tex
The deployment of PanDA Broker on Titan enabled distributed computing on a
leadership-class HPC machine at unprecedented scale. In the past 13 months,
PanDA WMS has consumed almost 52M core-hours on Titan, simulating 3.5\% of
the total number of detector events of the ATLAS production Monte Carlo
workflow. We described the implementation and execution process of PanDA WMS
(\S\ref{sec:panda_overview}) and PanDA Broker (\S\ref{sec:panda_titan}),
showing how they support and enable distributed computing at this scale on
Titan, a leadership-class HPC machine managed by OCLF.

We characterized the efficiency, scalability and reliability of both PanDA
Broker and AthenaMP as deployed on Titan (\S\ref{sec:panda_titan}). Our
characterization highlighted the strengths and limitations of the current
design and implementation: PanDA Brokers enable the sustained execution of
millions of simulations per week but further work is required to optimize its
efficiency and reliability (\S\ref{ssec:broker_titan}). PanDA Brokers support
the concurrent execution of multiple AthenaMP instances, enabling each
AthenaMP to perform the concurrent execution of up to 16 Geant4 simulators.
Nonetheless, our characterization showed how improving I/O performance could
reduce overheads (\S\ref{ssec:athenamp_titan}), increasing the overall
utilization of Titan's backfill availability. We introduced PanDA's next
generation executor for HPC systems, characterized its ability to support
multi-generation workloads and analyzed its scaling behavior.  Performance
tests on the next generation executor demonstrate linear strong and weak
scalability over several orders of magnitude  of task and node counts.
Qualitatively, it enables the high-performance execution of new workloads and
advanced execution modes of traditional workloads.

HEP was amongst the first, if not the first experimental community to realize
the importance of using WMS to manage their computational campaign(s). As
computing becomes increasingly critical for a range of experiments, the
experience foreshadows the importance of WMS for other experiments (such as
SKA, LSST etc.).  These experiments will have their own workload
characteristics, resources types and federation constraints, as well metrics
of performance. The experience captured in this paper should be useful for
designing WMS for computational campaigns and will provide a baseline to
evaluate the relative merits of different approaches.

The 52M core hours used by ATLAS, via PanDA, is over 2\% of the total
utilization on Titan over the same period, bringing the time-averaged
utilization of Titan to be consistently upwards of 90\%. Given that the
average utilization of most other leadership class machines is less (e.g.,
NSF's flagship Blue Waters the average utilization fluctuates between
60--80\% (see XDMoD\cite{bw-sucks})) there is ample headroom for similar
approaches elsewhere. These unprecedented efficiency gains aside, this work
is just a starting point towards more effective operational models for future
leadership and online analytical platforms~\cite{foap-url}. These platforms
will have to support ever increasing complex workloads with varying models
for dynamic resource federation.

{\bf \footnotesize Acknowledgements: This work is funded by Award number
DE-SC0016280 from the Office of Advanced Scientific Computing Research within
the Department of Energy. This research used resources of the Oak Ridge
Leadership Computing Facility at the Oak Ridge National Laboratory, which is
supported by the Office of Science of the U.S. Department of Energy under
Contract No. DE-AC05-00OR22725. \par}